\documentclass{jetpl}

\usepackage[cp1251]{inputenc}
\usepackage[english,russian]{babel}

\usepackage{mathtools}
\usepackage{xcolor}
\twocolumn

\rus

\newcommand{\ForceFigNubmer}[1]{
	Рис. \ref{#1}. 
	}

\graphicspath{{pictures/}}

\title{Построение структуры одномерного фотонного кристалла по заданному спектру коэффициента отражения}
\rtitle{Построение структуры фотонного кристалла...}
\sodtitle{Построение структуры одномерного фотонного кристалла по заданному спектру коэффициента отражения}
\author{П.\,С.\, Емельянцев$^+$, Н.\,И.\, Пышков$^+$, С.\,Е.\,Свяховский$^+$}
\rauthor{Емельянцев, Пышков, Свяховский}
\sodauthor{П.\,С.\, Емельянцев$^+$, С.\,Е.\,Свяховский$^+$}
\address{~\\$^+$Физический факультет МГУ имени~М.\,В.~Ломоносова, Москва, Ленинские горы, 1, стр. 62, 119991, Россия\\~}
\dates{25 апреля 2023 г.}{01 мая 2023 г.}
\abstract{Предложен и экспериментально реализован метод решения обратной задачи проектирования структуры одномерного фотонного кристалла. Известно, что одномерный фотонный кристалл, показатель преломления которого модулирован в виде синусоиды, имеет узкую фотонную запрещенную зону на частоте, однозначно связанной с пространственной частотой этой синусоиды.  Мы предлагаем метод обратного проектировании одномерных фотонных кристаллов с произвольным заданным спектром отражения путем разложения этого спектра по элементарным фотонным запрещенным зонам и последующего их суммирования. В работе показано 
применение этого метода для изготовления примеров фотонных кристаллов с простыми формами спектральных кривых отражения.}

\begin{document}

\maketitle

\textbf{Введение.} 
Интерес к структурам с периодически меняющимся показателем преломления берет свое начало с работ Яблоновича \cite{Yablonovitch} и Джона \cite{John}, в которых впервые было показано, что с помощью этих структур можно эффективно управлять распространением света аналогично тому, как полупроводники изменяют свойства электронов в физике твердого тела. Фотонные кристаллы (ФК) --- твердотельные структуры, в которых показатель преломления пространственно модулирован с периодом порядка длины волны. Такое устройство структуры ФК ведет к возникновению фотонных запрещенных зон (ФЗЗ), областей частот, в которых распространение света внутри ФК невозможно. Наличие ФЗЗ делает возможным возникновение некоторых интересных оптических эффектов, таких как усиление Рамановского рассеяния\cite{ashurov2020photonic}, эффект Парселла \cite{noda2007spontaneous}, генерация второй и третьей оптической гармоники \cite{martorell1997second, martemyanov2004third} и других.
ФК обладают исключительной гибкостью в возможностях управления потоком света, достигаемой посредством использования широкого числа конструктивных степеней свободы этих структур. Благодаря этой гибкости становится возможным реализация компактных и высокоэффективных логических устройств на основе фотонных структур, что критически важно для дальнейшего развития фотонных и оптоэлектронных технологий. В этом случае возникает проблема обратного проектирования ФК с заданными оптическими характеристиками, которая решалась различными способами, от методов математической оптимизации до применения самообучающихся алгоритмов.

\textbf{Существующие методы обратного проектирования ФК.} В работе \cite{Minkov} используется градиентная оптимизация - возможно наиболее популярная техника обратного проектирования фотонных структур \cite{Dop3, Dop4}. Задача оптимизации заключается в нахождении экстремумов целевой функции в некоторой области конечномерного векторного пространства. В данном случае аргументами такой функции являются настраиваемые параметры фотонного устройства --- показатели преломления и толщины слоёв. В рамках каждой итерации вычисляется градиент целевой функции по настраиваемым параметрам, а затем эти параметры меняются вдоль направления градиента для улучшения производительности фотонного устройства. Авторы работы \cite{Minkov} использовали метод градиентной оптимизации для улучшения добротности фотоннокристаллического резонатора малого объема более чем на два порядка. Также метод градиентной оптимизации может применяться и для обратного проектирования оптических \cite{Chen2013} и акустических \cite{Fahey} метаматериалов. Для обратного проектирования ФК используются такие методы оптимизации, как например, оптимизация на основе математической инверсии \cite{Geremia} или так называемая выпуклая оптимизация \cite{Convex}, которая в работе \cite{Piggott} была использована для создания компактного оптического демультиплексора.

Большая часть существующих работ в области решения задачи обратного проектирования фотонных структур в той или иной степени включает в себя использование машинного обучения, нейронных сетей (НС) и других подобных алгоритмов \cite{Dop5, Dop6, Dop7, Dop8}. В работе \cite{Deng} описывается метод, совмещающий классические алгоритмы оптимизации и НС для обратного проектирования ФК. Имея возможность изучать функции из обучающего набора данных, нейронные сети могут выполнять как прямое прогнозирование, так и и обратное проектирование различных фотонных структур. Объединение традиционных методов обратного проектирования и оптимизации с НС позволяет ещё больше повысить эффективность, гибкость и возможности моделей. Помимо НС в настоящее время в различных областях науки и производства набирают популярность алгоритмы машинного обучения. Основным их отличием от НС является то, что они требуют вмешательства человека в процессе их обучения, а также то, что они нацелены на более простые задачи и умеют работать только со структурированным набором данных. В работах \cite{Nikulin, Liu, Dop9, Dop10} показано успешное применение алгоритмов машинного обучения для обратного проектирования различных фотонных структур. Стоит также отметить возможность применения генетических алгоритмов \cite{Sanchez} для обратного проектирования ФК. 

Однако, вышеприведённые методы  не являются регулярными, в каждом конкретном случае существование и единственность решения не очевидны.

В этой работе мы предлагаем метод решения обратной задачи восстановления структур одномерных фотонных кристаллов (ФК) по спектру отражения. Рассматриваемый метод позволяет получить структуру ФК с произвольным заданным спектром отражения, что открывает огромный простор для возможных применений. Произведены численные расчеты нескольких структур ФК для различных заданных спектров отражения, были изготовлены опытные образцы одномерных ФК.

\textbf{Прямая задача: расчет спектра коэффициента отражения.} Вначале укажем использованное нами решение прямой задачи построения спектра ФК по известной структуре. Для этого используется метод матриц распространения \cite{luce2022tmm}. Мы используем рекуррентный метод, который является переформулированным матричным методом и на практике отличается более быстрой скоростью счёта.
Метод основан на классическом способе суммирования многократно отраженных лучей. Используемый нами рекуррентный метод позволяет вычислить коэффициенты отражения и прохождения структуры, состоящей из $m+1$ слоев, если эти коэффициенты известны для $m$ слоев \cite{Krylova}. Пусть на слоистую структуру, показатели преломления и толщины которой для каждого $m$-го слоя равны соответственно $n_{m}$ и $d_{m}$, под углом $\theta$ падает монохроматическая волна вида:

\begin{equation}
	E(\mathbf{r},t)= E(\mathbf{r}) e^{i \omega t}
	\label{E}
\end{equation}

Далее рассмотрим только пространственное распространение волны вдоль $x$ - нормального направления к поверхностям слоев, $E(\mathbf{r}) = E(x)$. Тогда поле в слое $m$ представимо в виде:

\begin{equation}
	E_{m}(x) = A_{m} e^{i k_{0} \sigma_{m}x} + B_{m} e^{-i k_{0}\sigma_{m}x}
\end{equation}

Здесь коэффициент $\sigma_{m} = (n_{m}^{2} - \sin^{2}\theta_{m})^{1/2}$, $k_{0}=2\pi / \lambda$ - волновой вектор падающей волны, $\theta_{m}$ - угол распространения преломленной волны к оси $x$. Волны с амплитудами $A_{m}$ и $B_{m}$ распространяются, соответственно, вдоль и противоположно оси $x$. Эти амплитуды могут быть получены из граничных условий для электрического и магнитного полей и из формул Френеля. Коэффициенты отражения от $m$-го и $(m+1)$-го слоев, $r_{m+1} = \frac{B_{m}}{A_{m}}$ и $r_{m+1} = \frac{B_{m+1}}{A_{m+1}}$ связаны рекуррентным соотношением Паррета:
\begin{equation}
	r_{m}=\frac{r_{m+1} + W_{m}}{1+r_{m+1}W_{m}}e^{2ik_{0}\sigma_{m}d_{m}}
\end{equation}
Данное выражение было получено с учетом непрерывности $x$-компонент волновых векторов  на границе раздела $m$-го и $(m+1)$-го слоев, а также граничные условия для s- и p-поляризаций света. Здесь введено обозначение:
\begin{equation}
	W_{m} = \frac{\sigma_{m} - \sigma_{m+1}P_{m}}{\sigma_{m} + \sigma_{m+1}P_{m}},
\end{equation}
где $P_{m}$ - поляризационный фактор, равный $P_{m} = 1$ для $S$-поляризации и $P_{m} = (\frac{n_{m}}{n_{m+1}})^{2}$ для $P$-поляризации. Расчет по рекуррентной формуле Паррета начинается с последнего слоя, для которого $r_{m+1}=0$. Коэффициент отражения для всей системы по интенсивности равен $R=r_{0}^{2}$. 

\begin{figure}
	\centering
	\includegraphics[width=\linewidth]{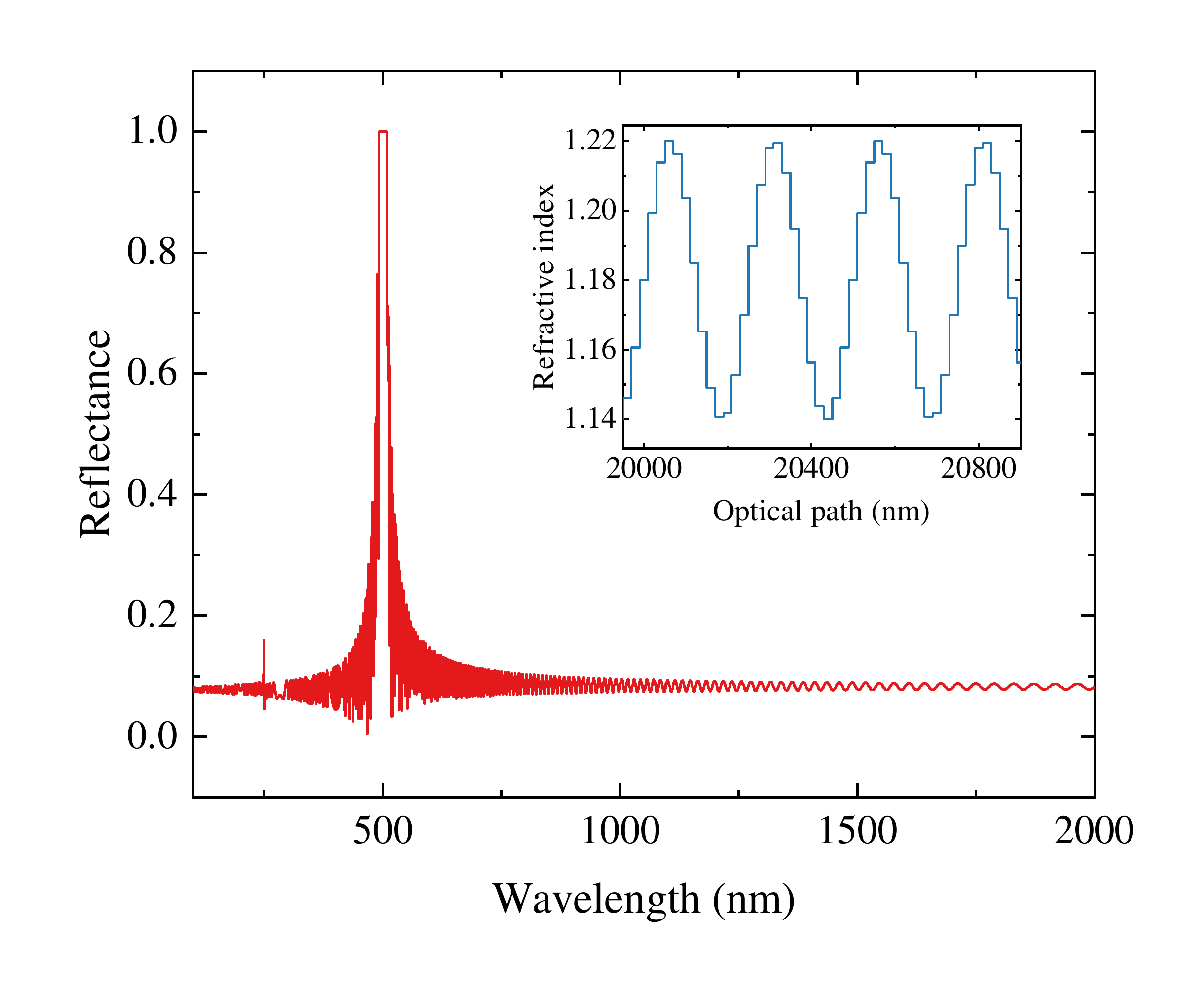}
	\caption{\ForceFigNubmer{delta} Теоретически рассчитанный спектр отражения ФК с модуляцией показателя преломления (\ref{sin1}) для $s$-поляризации. Результат получен для $L=40$ мкм, оптическом пути каждого слоя $\delta l =20$ нм, $n_{1}$ = 1.14, $n_{2}$ = 1.22, $\lambda_{0}$ = 500 нм. Вставка: зависимость показателя преломления от глубины фотонного кристалла, показан участок от 20.0 до 20.8 мкм.}
	\label{delta}
\end{figure}

\textbf{Обратная задача: метод построения структуры ФК.} Рассмотрим одномерный ФК суммарной оптической толщиной $L$, показатель преломления которого модулирован следующим образом: 
\begin{equation}
	n(x) = A \sin(k_{0}x + \phi) + B,
\end{equation}
где $x$ - оптический путь, $k_{0} = \frac{4\pi}{\lambda_{0}}$ - волновое число, $\phi$ - начальная фаза. Выберем $A$ и $B$ таким образом, чтобы $n(x)$ менялся в фиксированных пределах от $n_{1}$ до $n_{2} > n_{1}$: $A = (n_{2} - n_{1})$, $B = n_{1}$. В этом примере значение $\phi$ не повлияет на результат, поэтому выберем $\phi = 0$. Получаем:
\begin{equation}
	n(x) =  (n_{2} - n_{1}) \sin(k_{0}x) + n_{1}
	\label{sin1}
\end{equation}
ФК с такой модуляцией показателя преломления имеет очень узкую ФЗЗ на длине волны $\lambda_{0}$. Для вычисления спектра коэффициента отражения этой структуры приблизим непрерывную зависимость \ref{sin1} кусочно-постоянной, т.е. разделим кристалл на слои оптической толщиной $\delta l \ll \lambda_0$, Координата границы каждого $m$-го слоя  $x_m=m\cdot \delta l$, его показатель преломления $n_m=n(x_m)$. Физическая толщина слоёв определяется как $d_m=\delta l/n_m$. Приемлемость аппроксимации непрерывной функции $n(x)$ кусочно-постоянной обсуждается в \cite{baumeister1986simulation}. Результат вычисления показан  на рис. \ref{delta}. На вставке показана кусочно-постоянная зависимость показателя преломления от толщины кристалла, использованная при расчёте. В спектре действительно присутствует узкая ФЗЗ на заданной длине волны.

Это приводит нас к идее обратного проектирования одномерных ФК путем аппроксимации любого заданного спектра отражения этими узкими запрещёнными зонами. Естественное ограничение, накладываемое на задаваемый спектр, состоит в том, что он не должен иметь слишком узких спектральных особенностей: спектральная ширина особенности не должна быть меньше минимальной ширины запрещённой зоны.

Пусть $A(\lambda)$ - спектр отражения, для которого мы хотим подобрать соответствующую структуру одномерного ФК. Сделаем эту функцию дискретной: пусть этот спектр состоит из $N$ точек $a_{i}(\lambda_{i})$, $i\in[1,N]$. Пусть кристалл имеет $M$ слоев, толщина $m$-ого слоя $d_{m}$, оптический путь на каждом $m$-ом слое равен $l_{m} = n_{m}d_{m}$, $x_{m} = \sum\limits_{k=1}^{m}n_{m}d_{m}$ - оптический путь от поверхности ФК до начала ($m$+1)-ого слоя, $m\in [1,M]$. Показатель преломления меняется от $n_{1}$ до $n_{2}$. Тогда, в соответствии с уравнением (\ref{sin1}), получим выражение для показателя преломления $m$-ого слоя:
\begin{equation}
n(x_{m}) = \Delta n \left[ C_{1} \sum\limits_{i=1}^{N} a_{i}(\lambda_{i}) \sin \left( \frac{4\pi}{\lambda_{i}} x_{m} + \phi_{i} \right) + C_{2}\right] + n_{1}
\label{main_eq}
\end{equation}
Здесь  $C_{1}$, $C_{2}$ - нормировочные коэффициенты, которые выбраны таким образом, чтобы выражение в квадратных скобках попадало в интервал $[0,1]$. $\Delta n = n_2-n_1$. Так например, пусть для каждого оптического расстояния $x_{m}$ сумма $N$ гармонических функций равна: 
\begin{equation}
	S_{m} = \sum\limits_{i=1}^{N} a_{i}(\lambda_{i}) \sin \left( \frac{4\pi}{\lambda_{i}} x_{m} + \phi_{i} \right),
\end{equation}
и $S_{m}^{max}$, $S_{m}^{min}$ - максимальное и минимальное значения $S_{m}$ на протяжении всего кристалла соответственно, то $C_{1} = \frac{1}{S_{m}^{max} - S_{m}^{min}}$, $C_{2} = \frac{S_{m}^{min}}{S_{m}^{max} - S_{m}^{min}}$. Остается вопрос выбора фазы $\phi_i$. Для каждой $i$-ой гармонической функции фазовый множитель следующим образом: $\phi_{i} = 2\pi\frac{i}{N} \frac{L}{\lambda_m}$, где $\lambda_m$ - средняя длина волны спектрального диапазона, в котором решается обратная задача. Такая зависимость была выбрана, чтобы равномерно распределить по толщине кристалла возникающие между близкими гармониками биения.

\textbf{Примеры решения обратной задачи.} Для использования описанного выше метода в случае каждого заданного спектра необходимо выбрать суммарную толщину кристалла $L$. Далее кристалл разбивается на $M = L/\delta l$ слоёв, толщина каждого слоя $\delta l$ много меньше длины волны света. В данной работе выбрана величина $\delta l = 20$ нм.

\begin{figure}
	\centering
	\includegraphics[width=\linewidth]{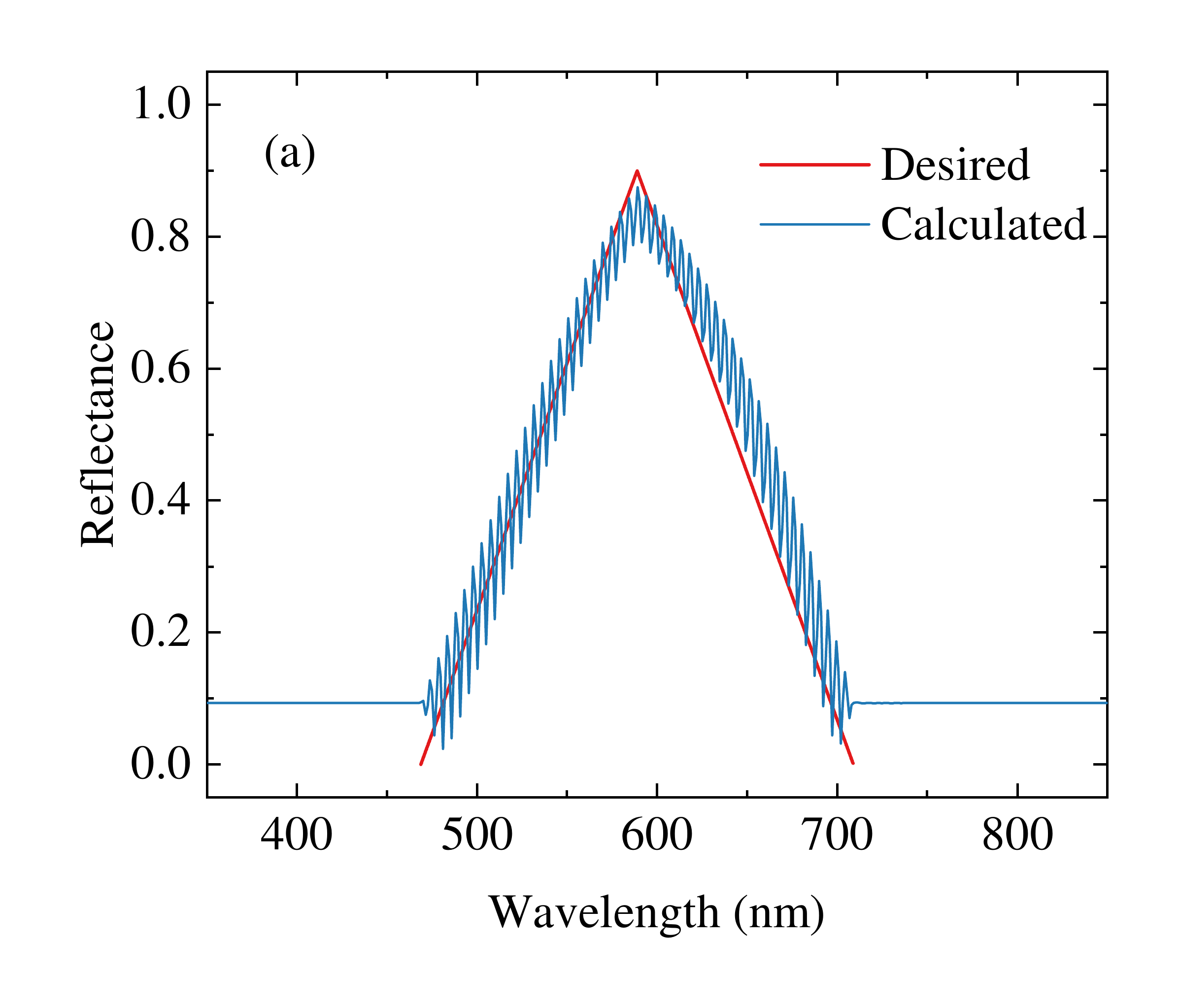}
	\includegraphics[width=\linewidth]{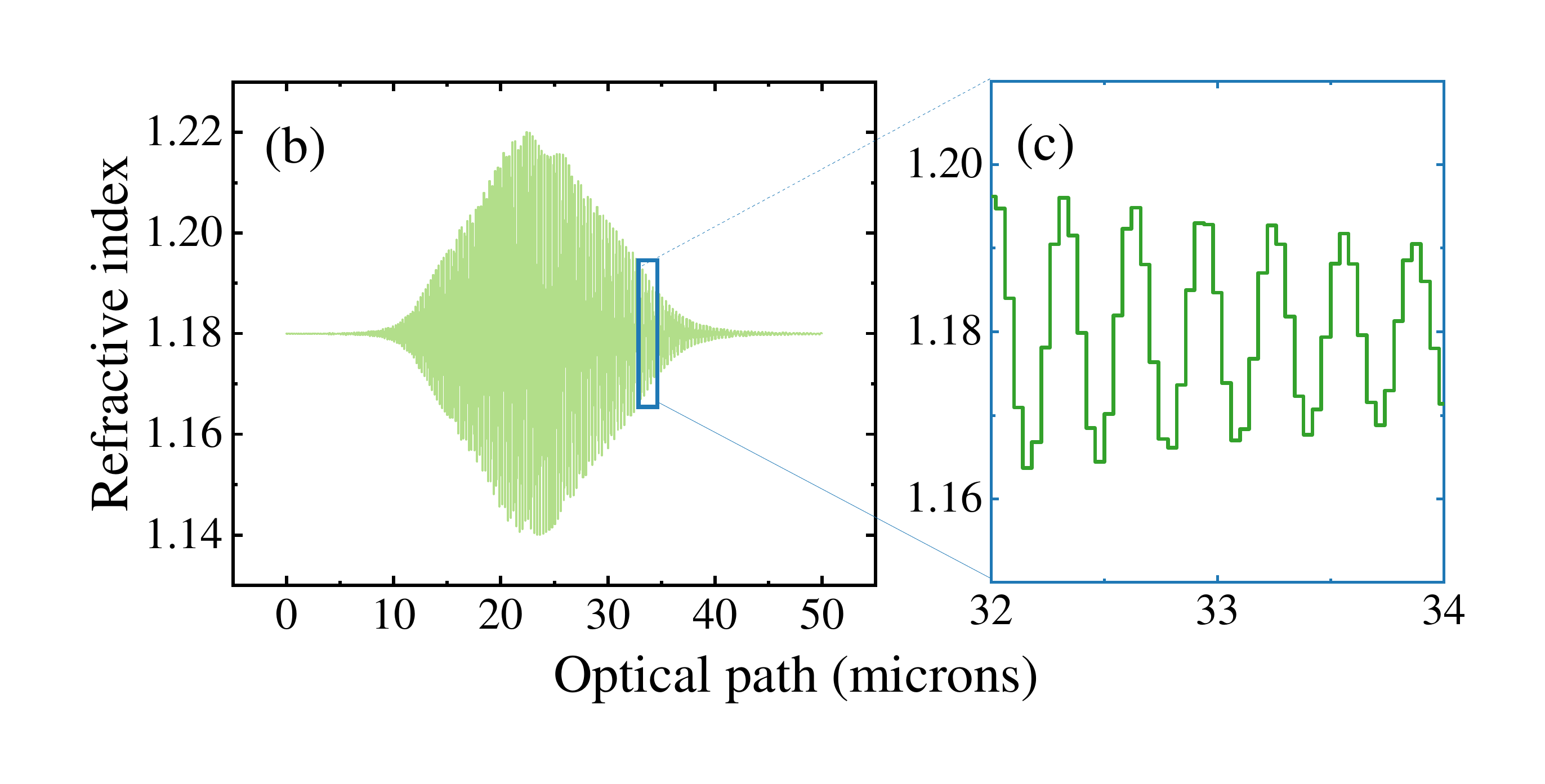}
	
	\caption{\ForceFigNubmer{Tr1} (цветной онлайн).Теоретические расчеты для фотонного кристалла со спектром отражения в виде треугольника: (a) задаваемый спектр отражения и спектр отражения, рассчитанный с помощью рекуррентного метода; (b) представление структуры данного кристалла как зависимость показателя преломления от оптического пути внутри кристалла, (с) увеличенный участок от 32 до 34 микрон.}
	\label{Tr1}
\end{figure}

В качестве первого примера была смоделирована структура одномерного ФК, спектр отражения которого представляет собой треугольную функцию шириной от 470 до 710 нм с вершиной на 590 нм, значение функции в вершине 0.9 --- такой выбор обусловлен тем, что функция отражения не может превосходить 1, и при этом единичное значение коэффициента отражения фотонного кристалла может быть получено слишком легко и тривиально.  Суммарный оптический путь $L$=50000 нм, количество слоев $M$ = 2500, число гармонических функций $N=256$. В качестве предельных значений показателей преломления слоев были выбраны  $n_{1}$ = 1.14 и $n_{2}$ = 1.22. Сравнение задаваемого спектра отражения и спектра отражения, рассчитанного рекуррентным методом для данной структуры ФК представлено на Рис. \ref{Tr1}(a), а вид данной структуры показан на Рис. \ref{Tr1}(b).  Спектр достаточно близок к желаемому, имеются паразитные осцилляции, вызванные интерференцией Фабри-Перо на границах структуры.

\begin{figure}
	\centering
	\includegraphics[width=0.9\linewidth]{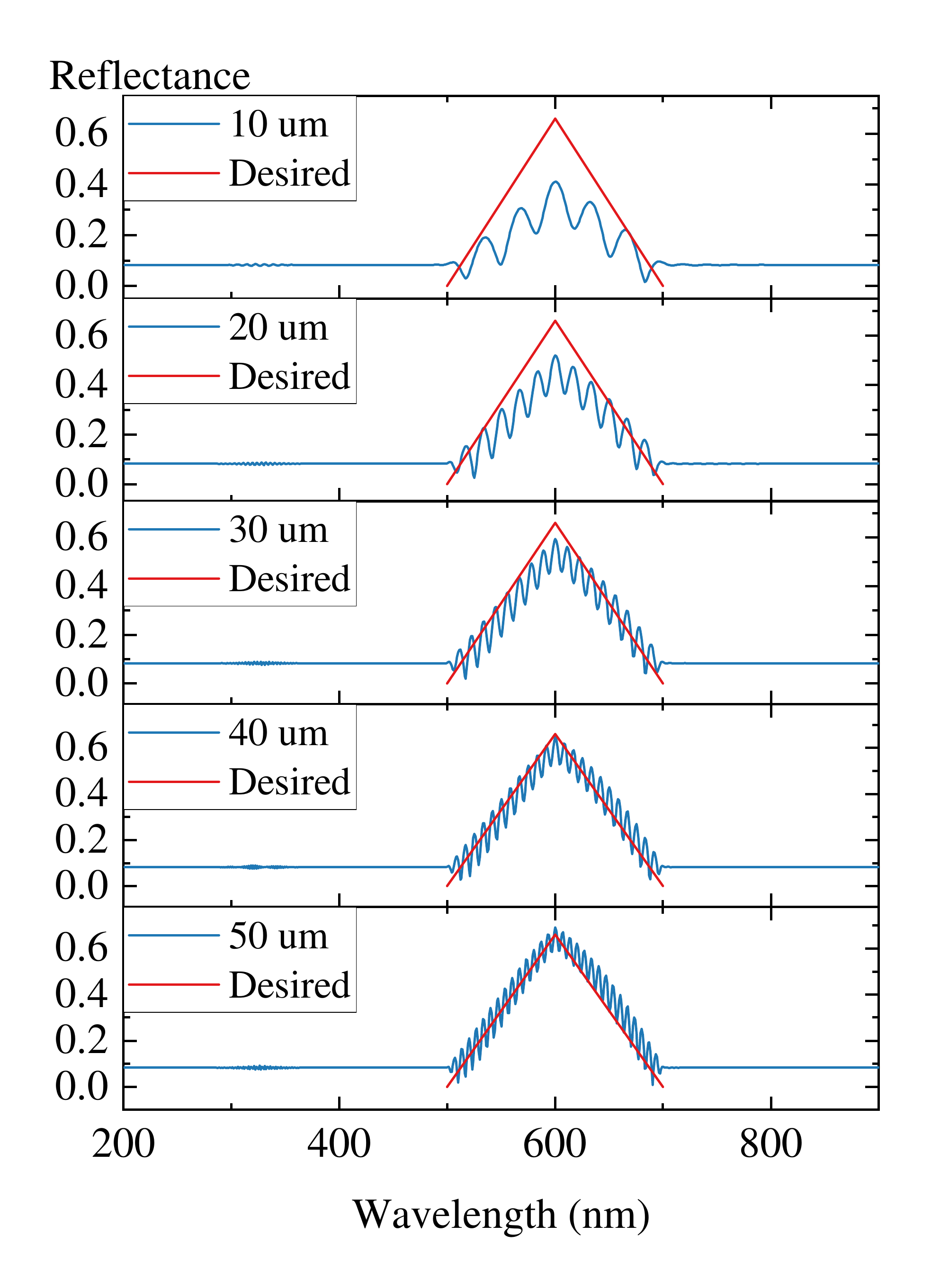}
	\caption{\ForceFigNubmer{Triangle5Spectra} (цветной онлайн). Теоретический расчёт фотонного кристалла с треугольным спектром отражения в сравнении с желаемым спектром для значений суммарной оптической толщины структуры $L$ от 10 до 50 мкм.}
	\label{Triangle5Spectra}
	
\end{figure}

Для выяснения значимости такого параметра, как толщина структуры, была смоделирована серия фотонных кристаллов с треугольной формой спектрального отклика. В пределах серии ставилось различное ограничение на суммарную оптическую толщину структуры $L$ от 10 до 50 мкм. Результаты моделирования  представлены на рис. \ref{Triangle5Spectra}. Заметно, что при увеличении толщины структуры уменьшается амплитуда и увеличивается частота паразитных осцилляций, спектр постепенно приближается к желаемому. Отметим, что число гармонических функций, которые используются для приближения спектра, во всех случаях постоянно и равно $N=256$, таким образом, их число не влияет на паразитные осцилляции. Отметим, что на длине волны 300 нм имеется незначительный артефакт - ФЗЗ второго порядка.

\begin{figure}
	\centering
	\includegraphics[width=\linewidth]{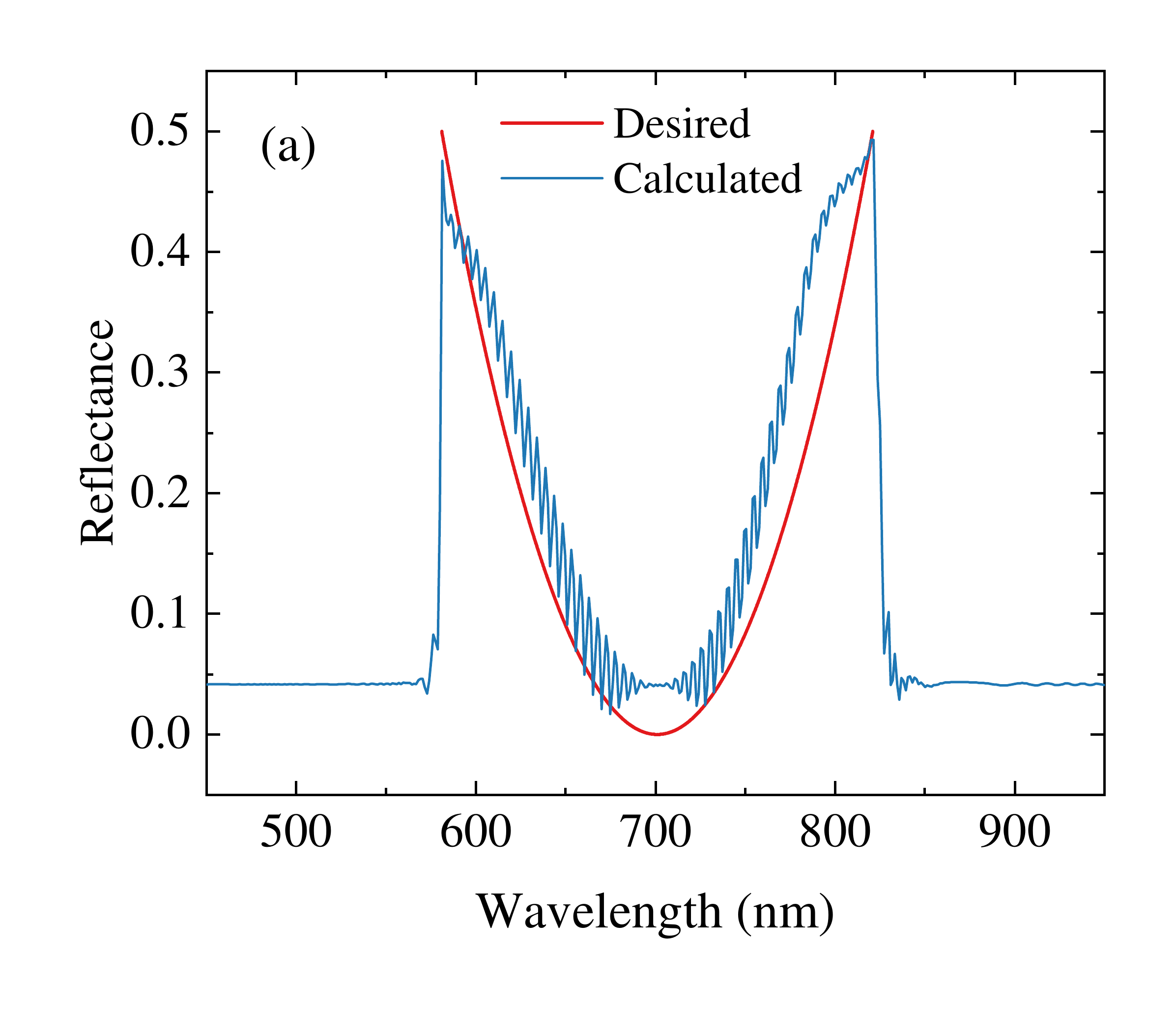}
	\includegraphics[width=\linewidth]{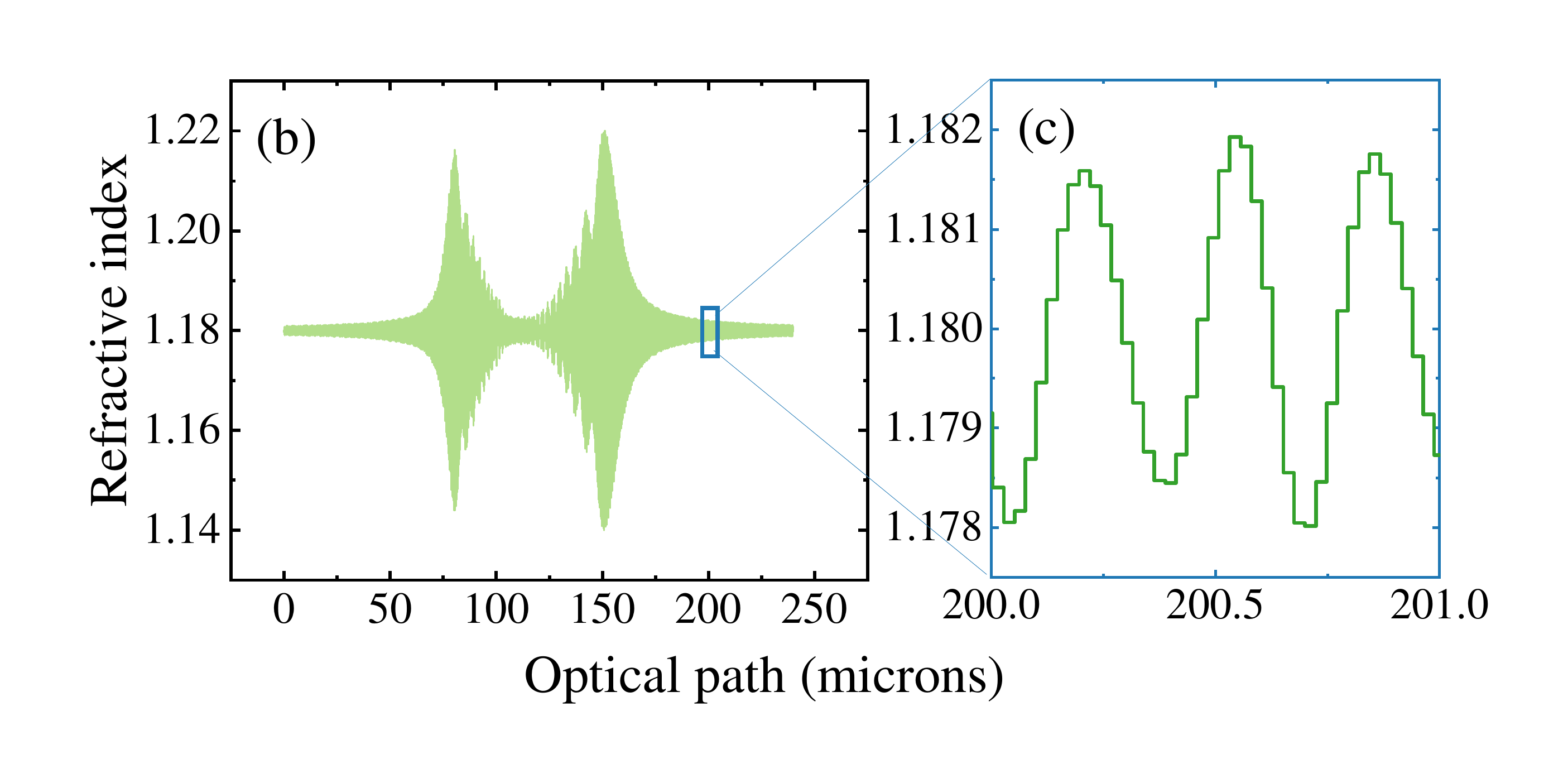}
	\caption{\ForceFigNubmer{Parab1}  (цветной онлайн). Теоретические расчеты для фотонного кристалла со спектром отражения в виде параболы: (a) задаваемый спектр отражения и спектр отражения, рассчитанный с помощью рекуррентного метода; (b) представление структуры данного кристалла как зависимость показателя преломления от оптического пути внутри кристалла, (с) увеличенный участок 200-201 мкм.}
	\label{Parab1}
\end{figure}

В качестве второго примера был смоделирован ФК со спектром отражения в виде параболы шириной от 580 до 820 нм с вершиной на 700 нм. Суммарный оптический путь $L$ = 250000 нм, предельные значения показателя преломления те же, что и для ФК с треугольным спектром. Сравнение задаваемого и рассчитанного спектров представлены на Рис. \ref{Parab1}(a), вид структуры ФК показан на Рис. \ref{Parab1}(b). Теоретический спектр достаточно хорошо аппроксимирует заданный. 

\textbf{Изготовление экспериментальных образцов.} Для экспериментальной проверки предлагаемого метода были изготовлены одномерные фотоннокристаллические структуры и измерены их оптические спектры. В данной работе для изготовления одномерных ФК используется методика электрохимического травления кремния, описанная в \cite{svyakhovskiy2012mesoporous}. Было показано, что при помощи данной методики можно изготовлять фотонные кристаллы с тысячами слоёв, при этом оптические потери определяются главным образом рэлеевским рассеянием, не превосходят нескольких процентов для середины оптического диапазона даже для образцов толщиной 100 мкм и быстро падают с ростом длины волны.

В этой работе использовались следующие параметры процесса травления: сырьём является кристаллический кремний ориентации поверхности (100), удельным сопротивлением 0.005 Ом$\cdot$см, минимальная и максимальная плотности токов $j_{min}=40$ мА/см$^2$, $j_{max}=160$ мА/см$^2$, в качестве электролита используется водно-спиртовой раствор фтороводородной кислоты (HF) в массовой концентрации 28\%.

\begin{figure}
	\centering
	\includegraphics[width=\linewidth]{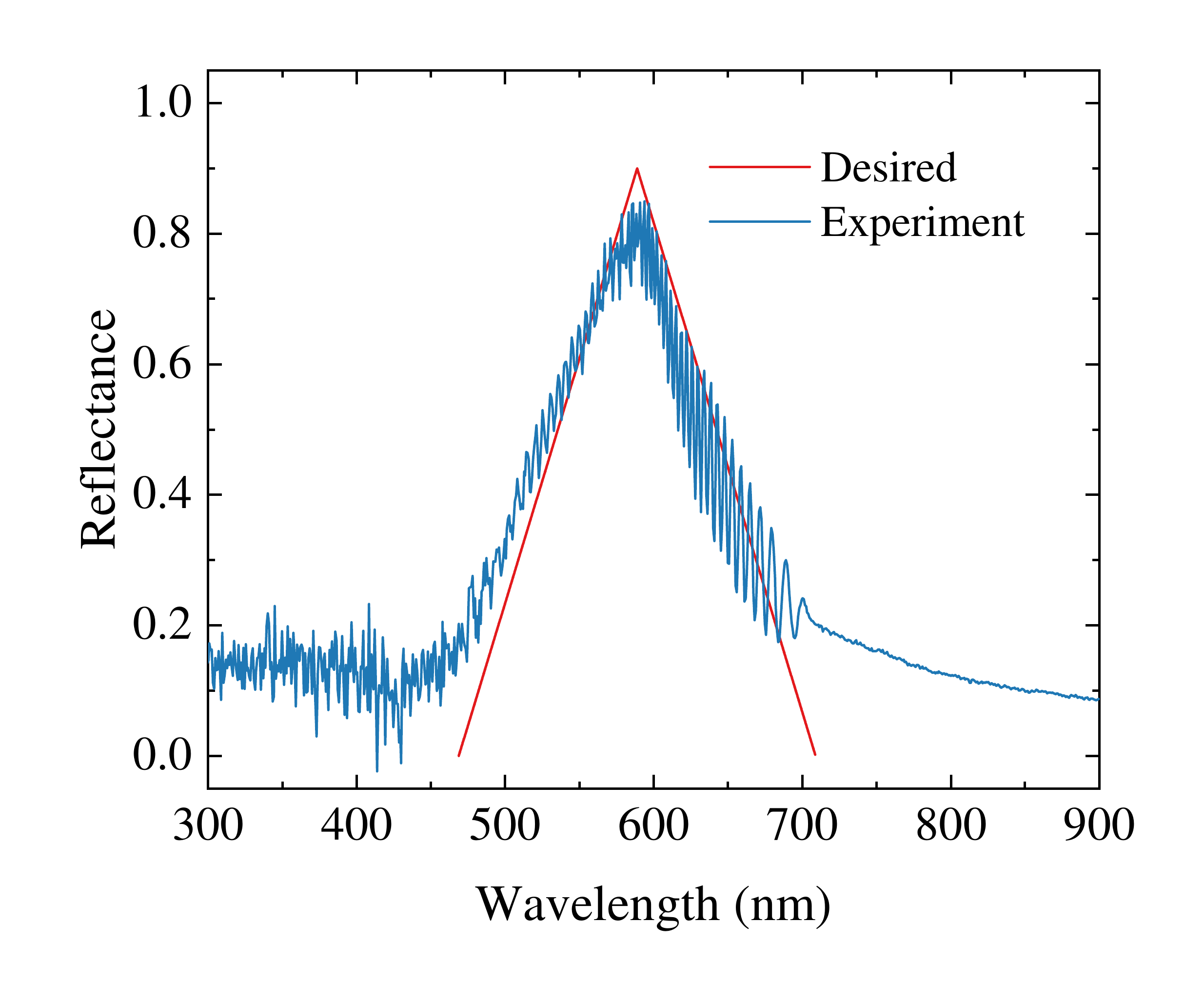}
	\caption{\ForceFigNubmer{Tr2}Экспериментальный спектр коэффициента отражения ФК в сравнении с желаемой функцией спектрального отклика, имеющей форму треугольника. }
	\label{Tr2}
\end{figure} 

\begin{figure}
	\centering
	\includegraphics[width=\linewidth]{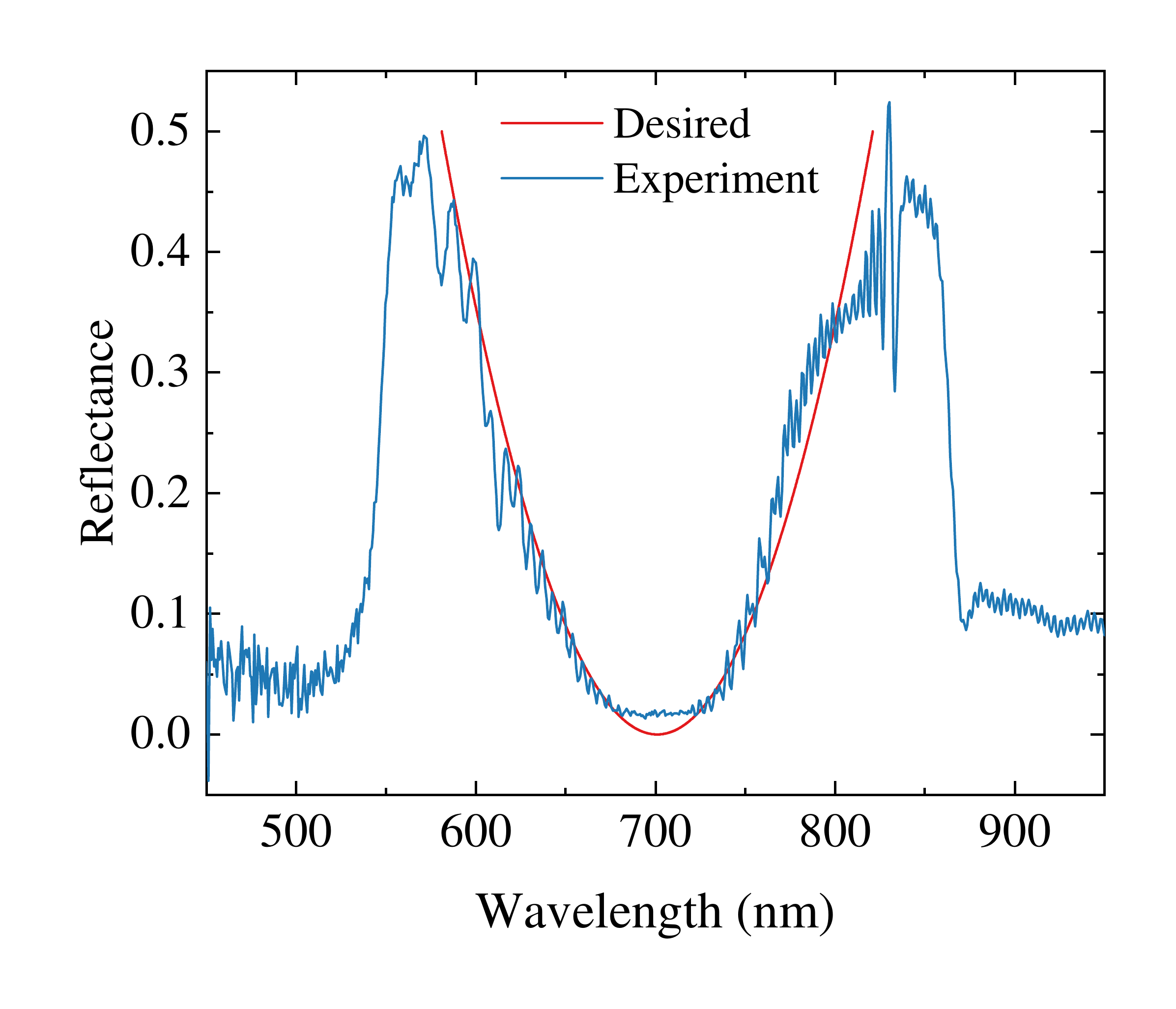}
	\caption{\ForceFigNubmer{Parab2}Экспериментальный спектр коэффициента отражения ФК в сравнении с желаемой функцией спектрального отклика, имеющей форму параболы.}
	\label{Parab2}
\end{figure}

На основе полученных расчетов были  изготовлены опытные образцы одномерных ФК, их модуляция показателя преломления соответствовала изображённым на рис. \ref{Tr1}(b) и \ref{Parab1}(b). Параметры образцов (профиль $n(x)$ и толщина $L$) соответствовали расчётным.
Измерение оптических спектров выполнялось по методике, описанной в \cite{svyakhovskiy2012mesoporous}. Спектры отражения полученных образцов снимались с помощью лабораторного спектрометра OceanInsight QEPRO при нормальном падении. Источником света служила галогеновая лампа со спектральным диапазоном 400-1200 нм. Измеренные спектры отражения для обоих примеров представлены на Рис. \ref{Tr2},\ref{Parab2}. Как можно увидеть, экспериментальные спектры находятся в хорошем соответствии с изначально заданными спектрами, как по форме, так и по положению. Так же, как и на теоретических графиках, присутствуют паразитные осцилляции, амплитуда которых не превышает 10\% от величины полезного сигнала. Для случая параболического спектра на краях диапазона имеются уширения, связанные с конечной минимальной шириной запрещённой зоны ФК. Экспериментально полученная форма функции в заданной области соответствует заданной.

\textbf{Заключение.} Был продемонстрирован новый метод обратного проектирования одномерных ФК, позволяющий по заданной спектральной функции коэффициента отражения построить структуру ФК как зависимость показателя преломления от глубины структуры. Показаны примеры применения метода на спектральных функциях простой формы. Метод подтверждён экспериментально, образцы ФК были изготовлены при помощи электрохимического травления кремния. Рассматриваемый метод может быть пригоден для  других методов изготовления, в которых имеется техническая возможность задавать произвольный пространственный профиль $n(x)$, например, электрохимическое травление алюминия, титана и двухфотонная фотополимеризация с градиентным изменением показателя преломления.

Авторы выражают благодарность за финансовую поддержку этой работы Российским научным фондом, проект 21-72-10103, https://rscf.ru/en/project/21-72-10103/.

\bibliographystyle{sse-jetpl}
\bibliography{bibliography}
\end{document}